\documentstyle[11pt,times]{article}
\mathcode`\0="0030      %
\mathcode`\1="0031
\mathcode`\2="0032
\mathcode`\3="0033
\mathcode`\4="0034
\mathcode`\5="0035
\mathcode`\6="0036
\mathcode`\7="0037
\mathcode`\8="0038
\mathcode`\9="0039

\makeatletter
\def\@begintheorem#1#2{\trivlist\item[\hskip\labelsep{\bf #1\ #2}]}
\def\foobarpt{\textfont\z@\tenrm 
  \scriptfont\z@\ninrm \scriptscriptfont\z@\sevrm
\textfont\@ne\tenmi \scriptfont\@ne\ninmi \scriptscriptfont\@ne\sevmi
\textfont\tw@\tensy \scriptfont\tw@\ninsy \scriptscriptfont\tw@\sevsy
\textfont\thr@@\tenex \scriptfont\thr@@\tenex \scriptscriptfont\thr@@\tenex
\def\unboldmath{\everymath{}\everydisplay{}\@nomath\unboldmath
          \textfont\@ne\tenmi 
          \textfont\tw@\tensy \textfont\lyfam\tenly
          \@boldfalse}\@boldfalse
\def\boldmath{\@ifundefined{tenmib}{\global\font\tenmib\@mbi\@magscale1\global
        \font\tensyb\@mbsy \@magscale1\global\font
         \tenlyb\@lasyb\@magscale1\relax\@addfontinfo\@xiipt
              {\def\boldmath{\everymath
                {\mit}\everydisplay{\mit}\@prtct\@nomathbold
                \textfont\@ne\tenmib \textfont\tw@\tensyb 
                \textfont\lyfam\tenlyb\@prtct\@boldtrue}}}{}\@xiipt\boldmath}%
\def\prm{\fam\z@\tenrm}%
\def\pit{\fam\itfam\tenit}\textfont\itfam\tenit \scriptfont\itfam\ninit
   \scriptscriptfont\itfam\sevit
\def\psl{\fam\slfam\tensl}\textfont\slfam\tensl 
     \scriptfont\slfam\tensl \scriptscriptfont\slfam\tensl
\def\pbf{\fam\bffam\tenbf}\textfont\bffam\tenbf 
   \scriptfont\bffam\ninbf \scriptscriptfont\bffam\ninbf 
\def\ptt{\fam\ttfam\tentt}\textfont\ttfam\tentt
   \scriptfont\ttfam\nintt \scriptscriptfont\ttfam\nintt 
\def\psf{\fam\sffam\tensf}\textfont\sffam\tensf
    \scriptfont\sffam\tensf \scriptscriptfont\sffam\tensf
\def\psc{\@getfont\psc\scfam\@xiipt{\@mcsc\@magscale1}}%
\def\ly{\fam\lyfam\tenly}\textfont\lyfam\tenly 
   \scriptfont\lyfam\ninly \scriptscriptfont\lyfam\sevly
 \@setstrut \rm}

\newcommand{\singlespacing}{\let\CS=
\@currsize\renewcommand{\baselinestretch}{1}\tiny\CS}
\newcommand{\singlespacingplus}{\let\CS=
\@currsize\renewcommand{\baselinestretch}{1.25}\tiny\CS}
\newcommand{\doublespacing}{\let\CS=
\@currsize\renewcommand{\baselinestretch}{1.75}\tiny\CS}
\newcommand{\draftspacing}{\let\CS=
\@currsize\renewcommand{\baselinestretch}{2.0}\tiny\CS}

\newcommand{\niceonespacing}{\let\CS=\@currsize\renewcommand{\baselinestretch}{1.1}\tiny\CS}\newcommand{\nicetwospacing}{\let\CS=\@currsize\renewcommand{\baselinestretch}{1.2}\tiny\CS}
\newcommand{\nicethreespacing}{\let\CS=\@currsize\renewcommand{\baselinestretch}{1.3}\tiny\CS}
\newcommand{\singlespacingplusplus}{\let\CS=\@currsize\renewcommand{\baselinestretch}{1.35}\tiny\CS}
\newcommand{\nicefivespacing}{\let\CS=\@currsize\renewcommand{\baselinestretch}{1.5}\tiny\CS}
\newcommand{\nicesixpacing}{\let\CS=\@currsize\renewcommand{\baselinestretch}{1.6}\tiny\CS}

\makeatother

\makeatletter
\def\@cite#1#2{[#1\if@tempswa , #2\fi]}
\makeatother

\newcommand\seq{\subseteq}
\renewcommand\.{\cdot}
\newcommand\<{\langle}
\renewcommand\>{\rangle}

\newcommand\Lora{\ \Longrightarrow \ }
\newcommand\Lolra{\ \Longleftrightarrow \ }

\newcommand\equalsdef{\stackrel{\mbox{\protect\scriptsize df}}{=}}

\newcommand\tweak{\hspace*{1pt}}

\newcommand\xor{\mbox{$\sym$}}
\newcommand\XOR{\mbox{\boldmath $\,\sym\,$}}

\newcommand\union{\mbox{\boldmath $\,\vee\,$}}
\newcommand\inter{\mbox{\boldmath $\,\wedge\,$}}
\newcommand\minus{\mbox{\boldmath $\, -\,$}}

\def\sym{\Delta}

\newcommand\p{\mbox{\rm P}}

\newcommand\np{\mbox{\rm NP}}
\newcommand\ph{\mbox{\rm PH}}
\newcommand\conp{\mbox{\rm coNP}}
\newcommand\coup{\mbox{\rm coUP}}
\newcommand\co{\mbox{\rm co}}

\newcommand\up{{\mbox{UP}}}
\newcommand\prup{{\mbox{\tweak$\cal UP$}}}
\renewcommand\u{\mbox{U}}
\newcommand\pru{{\mbox{\tweak$\cal U$}}}
\newcommand\pruph{{\mbox{\tweak$\cal UPH$}}}

\renewcommand\c[1]{\mbox{\rm C}_{#1}}
\newcommand\ch{{\mbox{CH}}}
\renewcommand\d[1]{\mbox{\rm D}_{#1}}
\newcommand\newdh{\mbox{DH}}
\newcommand\e[1]{\mbox{\rm E}_{#1}}
\newcommand\eh{\mbox{EH}}
\newcommand\sd[1]{\mbox{SD}_{#1}}
\newcommand\sdh{{\mbox{SDH}}}

\newcommand\dnp{\mbox{DP}}
\newcommand\bcup{\mbox{BC$(\up)$}}

\newcommand\smallprup{{\mbox{\tweak\scriptsize \prup}}}
\newcommand\smallpru{{\mbox{\tweak\scriptsize \pru}}}

\newcommand\tinypru{{\mbox{\tiny \pru}}}

\newcommand\sigp[1]{\Sigma^p_{#1}}
\newcommand\pip[1]{\Pi^p_{#1}}
\newcommand\delp[1]{\Delta^p_{#1}}

\newtheorem{theorem}{Theorem}[section]
\newtheorem{corollary}[theorem]{Corollary}

\newtheorem{fact}[theorem]{Fact}

\newtheorem{definition}[theorem]{Definition}

\newtheorem{thm}[theorem]{Theorem}
\newtheorem{lem}[theorem]{Lemma}

\newtheorem{cor}[theorem]{Corollary}
\newtheorem{remark}[theorem]{Remark}

\newenvironment{construction}{\bigbreak\begin{block}}{\end{block}
    \bigbreak}

\newenvironment{block}{\begin{list}{\hbox{}}{\leftmargin 1em
    \itemindent -1em \topsep 0pt \itemsep 0pt \partopsep 0pt}}{\end{list}}

\load{\scriptsize}{\cal}
\load{\tiny}{\cal}

\begin{document}

\bibliographystyle{alpha}

\title{
Unambiguous Computation:  Boolean Hierarchies and Sparse
Turing-Complete Sets}

\author{ 
{\em  Lane A. Hemaspaandra\/}\thanks{Supported 
in part by 
grants NSF-CCR-8957604, 
NSF-INT-9116781/\allowbreak{}JSPS-ENGR-207, and
NSF-CCR-9322513, and by an
NAS/NRC COBASE grant. Email: lane@cs.rochester.edu.} \\
Department of Computer Science \\
University of Rochester \\
Rochester, NY 14627
\and 
{\em  J\"org Rothe\/}\thanks{Supported 
in part by a grant
from the DAAD and by grant NSF-CCR-8957604. 
Work done in part while visiting the University of Rochester.
Email: rothe@mipool.uni-jena.de.} \\
Institut f\"ur Informatik \\
Friedrich-Schiller-Universit\"at Jena \\
07743 Jena, Germany
}

\newcount\hour  \newcount\minutes  \hour=\time  \divide\hour by 60
\minutes=\hour  \multiply\minutes by -60  \advance\minutes by \time
\def\mmmddyyyy{\ifcase\month\or Jan\or Feb\or Mar\or Apr\or May\or Jun\or Jul\or
  Aug\or Sep\or Oct\or Nov\or Dec\fi \space\number\day, \number\year}
\def\hhmm{\ifnum\hour<10 0\fi\number\hour :%
  \ifnum\minutes<10 0\fi\number\minutes}
\def\Draft{{\it Draft of \mmmddyyyy}}

\date{}

\typeout{WARNING:  BADNESS used to supress reporting!  Beware!!}
\hbadness=3000%
\vbadness=10000 %

\pagestyle{empty}
\setcounter{footnote}{0}
{\singlespacing\maketitle}

\begin{center}
{\large\bf Abstract}
\end{center}
\begin{quotation}
{\singlespacing

  \noindent 
  It is known that for any class $\tweak{\cal C}$ closed under {\em
    union and intersection\/}, the Boolean closure of ${\cal C}$, the
  Boolean hierarchy over $\tweak{\cal C}$, and the symmetric
  difference hierarchy over $\tweak{\cal C}$ all are equal.
  We prove that these equalities hold for any complexity class closed
  under {\em intersection\/}; in particular, they thus hold for
  unambiguous polynomial time (UP). In contrast to the NP case, we
  prove that the Hausdorff hierarchy and the nested difference
  hierarchy over UP both fail to capture the Boolean closure of UP in
  some relativized worlds.

  Karp and Lipton proved that if {\em nondeterministic\/} polynomial
  time has sparse Turing-complete sets, then the polynomial hierarchy
  collapses.
  We establish the first
  consequences from the assumption that {\em
  unambiguous\/} polynomial time has sparse Turing-complete sets:
(a)~$\up \seq \mbox{Low}_2$, where $\mbox{Low}_2$ is the second level 
of the low
hierarchy,
and (b)~each level of the unambiguous
polynomial hierarchy
is contained one
level lower in the promise unambiguous polynomial
hierarchy
than is otherwise known to be the case.
}
\end{quotation}

\nicetwospacing
\setcounter{page}{1}
\pagestyle{plain}

\sloppy

\section{Introduction}

NP and NP-based hierarchies---such as the polynomial
hierarchy~\cite{mey-sto:b:reg-exp-needs-exp-space,sto:j:poly} and the
Boolean hierarchy over
NP~\cite{cai-gun-har-hem-sew-wag-wec:j:bh1,cai-gun-har-hem-sew-wag-wec:j:bh2,koe-sch-wag:j:diff}---have
played such a central role in complexity theory, and have been so
thoroughly investigated, that it would be natural to take them as
predictors of the behavior of other classes or hierarchies.  However,
over and over during the past decade it has been shown that NP is a
singularly poor predictor of the behavior of other classes (and, to a
lesser extent, that hierarchies built on NP are poor predictors of
the behavior of other hierarchies).

As examples regarding hierarchies: though the polynomial hierarchy
possesses downward separation (that is, if its low levels collapse, then
all its levels collapse)~\cite{mey-sto:b:reg-exp-needs-exp-space,sto:j:poly}, 
downward 
separation does not hold ``robustly'' (i.e., in every relativized world)
for the exponential time hierarchy~\cite{har-imm-sew:j:sparse,imp-tar:c:dec}
or for limited-nondeterminism 
hierarchies~(\cite{hem-jha:cOUTbyJtoAppear:defying}, see
also~\cite{bei-gol:c:beta}).
\typeout{FILL IN THE JOURNAL VERSION OF THIS}
As examples regarding UP:  NP has $\leq_m^p$-complete sets, but 
UP does not robustly possess $\leq_m^p$-complete
sets~\cite{har-hem:j:up} or even
$\leq_T^p$-complete sets~\cite{hem-jai-ver:j:up-turing};
NP positively relativizes, in the sense that it collapses to
P if and only if it does so with respect to every tally 
oracle (\cite{lon-sel:j:sparse}, see also~\cite{bal-boo-sch:j:sparse}),
but UP does not robustly positively 
relativize~\cite{hem-rub:j:positive};
NP has ``constructive programming systems,''
but UP does not robustly have such systems~\cite{reg:unpub:cons};
NP (actually, nondeterministic computation) admits 
time hierarchy theorems~\cite{har-ste:j:algs}, but 
it is an open question whether unambiguous computation
has nontrivial time hierarchy theorems;  NP displays
upward separation (that is, $\mbox{NP} - \mbox{P}$ contains
sparse sets if and only if 
$\mbox{NE} \neq \mbox{E}$)~\cite{har-imm-sew:j:sparse},
but it is not known whether UP does 
(see~\cite{hem-jha:cOUTbyJtoAppear:defying}, which
shows that R and BPP do not robustly display upward separation,
and~\cite{rao-rot-wat:j:upward}, which shows that FewP
does possess upward separation).

In light of the above list of the many ways in which NP parts company
with UP, it is clear that we should not merely assume that results for
NP hold for UP, but, rather, we must carefully check to see to what
extent, if any, results for NP suggest results for UP\@.  In this
paper, we study, for UP, two topics that have been intensely studied
for the NP case: the structure of Boolean hierarchies, and the effects
of the existence of sparse Turing-complete/Turing-hard sets.

For the Boolean hierarchy over NP, which has generated quite a bit of
interest and the collapse of which is known to imply the collapse of
the polynomial
hierarchy~\cite{kad:joutdatedbychangkadin:bh,cha-kad:c:closer,bei-cha-ogi:j:difference-hierarchies},
a large number of definitions are known to be equivalent.  For
example, for NP, all the following
coincide~\cite{cai-gun-har-hem-sew-wag-wec:j:bh1}: the Boolean closure
of NP, the Boolean (alternating sums) hierarchy, the nested
difference hierarchy, and the 
Hausdorff hierarchy.  
The symmetric difference hierarchy also characterizes the
Boolean closure of NP~\cite{koe-sch-wag:j:diff}.  
In fact, these equalities are known to hold for 
all classes that contain $\Sigma^*$ and $\emptyset$
and are closed under union and 
intersection~\cite{hau:b:sets,cai-gun-har-hem-sew-wag-wec:j:bh1,koe-sch-wag:j:diff,ber-bru-jos-sit-you:c:gen,gun-nas-wec:c:counting-survey,cha-kad:t:characteristic,cha:thesis}.
In Section~\ref{section:BH}, we prove that both the symmetric 
difference hierarchy (SDH) and the Boolean hierarchy (CH)
remain equal to the Boolean closure (BC) {\em even in the 
absence of the assumption of closure under 
union}.  That is, for any class $\cal K$ containing $\Sigma^*$ and 
$\emptyset$ and closed under intersection (e.g.,
UP, US, and DP, first defined respectively
in~\cite{val:j:checking},~\cite{bla-gur:j:unique-sat},
and~\cite{pap-yan:j:dp} and each of which is not currently known to be closed
under union):
$\sdh({\cal K}) = \ch({\cal K}) = \mbox{BC}({\cal K})$.  However, for
the remaining two hierarchies, we show that not all classes containing
$\Sigma^*$ and $\emptyset$ and closed under intersection robustly
display equality.
In particular,
the Hausdorff hierarchy over UP and the nested difference hierarchy
over UP both fail to robustly capture the Boolean closure of UP\@.  In
fact, the failure is relatively severe; we show that even low levels
of other Boolean hierarchies over UP---the third level of the
symmetric difference hierarchy and the fourth level of the Boolean
(alternating sums) hierarchy---fail to be robustly captured by either
the Hausdorff hierarchy or the nested difference hierarchy.  

It is well-known, thanks to the work of Karp and
Lipton~(\cite{kar-lip:c:nonuniform}, see also the related references
given in Section~\ref{section:Turing-Complete-Sets}), that if NP has
sparse Turing-hard sets, then the polynomial hierarchy collapses.
Unfortunately, the promise-like definition of UP---its unambiguity,
the very core of its nature---seems to block any similarly strong
claim for UP and the unambiguous polynomial hierarchy (which was
introduced recently by Niedermeier and
Rossmanith~\cite{nie-ros:c:up-hierarchy}).
Section~\ref{section:Turing-Complete-Sets} studies this issue, and
shows that if UP has sparse Turing-complete sets, then the levels of
the unambiguous polynomial hierarchy ``slip down'' slightly in terms
of their location within the promise unambiguous polynomial hierarchy
(a version of the unambiguous polynomial hierarchy that requires only
that computations {\em actually executed\/} be unambiguous), i.e., the
$k$th level of the unambiguous polynomial hierarchy is contained in
the $(k-1)$st level of the promise unambiguous polynomial hierarchy.
Various related results are also established.  For example, if UP has
Turing-hard sparse sets, then (a) $\up \seq \mbox{Low}_2$,
where $\mbox{Low}_2$ is the second level of the 
low hierarchy~\cite{sch:j:low}, and (b)
the $k$th level of the unambiguous polynomial hierarchy can be
accepted via a deterministic polynomial-time Turing transducer given
access to both a ${\Sigma}_2^p$ set and the $(k-1)$st level of the
promise unambiguous polynomial hierarchy.

\section{Notations}

In general, we adopt the standard notations of Hopcroft and
Ullman~\cite{hop-ull:b:automata}.  Fix the alphabet $\Sigma =
\{0,1\}$. $\Sigma^*$ is the set of all strings over $\Sigma$.  For
each string $u\in \Sigma^{*}$, $|u|$ denotes the length of~$u$.
The empty string is denoted by $\epsilon$.
For each set $L\subseteq \Sigma^{*}$, $\|L\|$ denotes the cardinality
of $L$ and $\overline{L} = \Sigma^{*}-L$ denotes the complement of
$L$\@.  $L^{=n}$ ($L^{\leq n}$) is the set of all strings in $L$
having length~$n$ (less than or equal to $n$).  Let $\Sigma^n$ and
$\Sigma^{\leq n}$ be shorthands for $(\Sigma^*)^{=n}$ and
$(\Sigma^*)^{\leq n}$, respectively.  A set~$S$ is said to be {\em
  sparse\/} if there is a polynomial $q$ such that for every $m\geq
0$, $\|S^{\leq m}\|\leq q(m)$.  To encode a pair of strings, we use a
polynomial-time computable pairing function, $\<\.,\.\>:\Sigma^*\times
\Sigma^* \,\rightarrow\, \Sigma^*$, that has polynomial-time
computable inverses; this notion is extended to encode every $k$-tuple
of strings, in the standard way.  Let $\leq_{\mbox{\protect\scriptsize lex}}$
denote the standard quasi-lexicographical ordering on $\Sigma^{*}$,
that is, for strings $x$ and $y$, $x\leq_{\mbox{\protect\scriptsize lex}} y$
if either $x=y$, or $|x|<|y|$, or $(|x|=|y|$ and there exists some
$z\in \Sigma^{*}$ such that $x=z0u$ and $y=z1v)$.  $x
<_{\mbox{\protect\scriptsize lex}} y$ indicates that $x\leq_{\mbox{\protect\scriptsize
    lex}} y$ but $x\neq y$.

For sets $A$ and $B$, their join, $A\oplus B$, is $\{0x\,|\,x\in
A\}\cup\{1x\,|\,x\in B\}$, and their symmetric difference, $A\sym B$,
is $(A-B) \cup (B-A)$.  
For any class $\tweak{\cal C}$, define
$\co\tweak{\cal C} \equalsdef \{L\,|\,\overline{L}\in
{\cal C}\}$, and let $\mbox{BC}({\cal C})$ denote the Boolean algebra
generated by $\tweak{\cal C}$, i.e., the smallest class containing
$\tweak{\cal C}$ and closed under all Boolean operations.  
For any classes
$\tweak\cal A$ and $\tweak\cal B$, let
$\tweak{\cal A} \oplus \tweak{\cal B}$ denote the class 
\mbox{$\{ A \oplus B \,|\, A \in {\cal A} \,\wedge\, B \in {\cal B}\}$}.
Similarly, for classes $\tweak\cal C$ and $\tweak\cal D$ of sets, define
\[
\begin{array}{lcllcl}
{\cal C} \inter {\cal D} & \equalsdef & 
\{A \cap B \,|\, A\in {\cal C}\wedge B\in {\cal D}\}, &
{\cal C} \XOR {\cal D}   & \equalsdef &
\{A \,\sym\, B \,|\, A\in {\cal C}\wedge B\in {\cal D}\}, \\
{\cal C} \union {\cal D} & \equalsdef & 
\{A \cup B \,|\, A\in {\cal C}\wedge B\in {\cal D}\}, &
{\cal C} \minus {\cal D} & \equalsdef & 
\{A - B \,|\, A\in {\cal C}\wedge B\in {\cal D}\}. 
\end{array}
\]

We will abbreviate ``polynomial-time deterministic (nondeterministic)
Turing machine'' by DPM (NPM). An {\em unambiguous\/} (sometimes
called categorical) polynomial-time Turing machine (UPM) is an NPM
that on no input has more than one accepting computation
path~\cite{val:j:checking}.  \up\ is the class of all languages that
are accepted by some UPM~\cite{val:j:checking}.  For the respective
oracle machines we use the shorthands DPOM, NPOM, and UPOM.

Note, crucially, that whether a machine is categorical or not depends
on its oracle.  In fact, it is well-known that machines that are
categorical with respect to all oracles accept only easy
languages~\cite{har-hem:j:rob} and thus create a polynomial hierarchy
analog that is completely contained in a low level of the polynomial
hierarchy (Allender and Hemachandra as cited
in~\cite{hem-rub:j:positive}).  So, when we speak of a UPOM, we will
simply mean an NPOM that, with the oracle the machine has in the
context being discussed, happens to be categorical.

For any Turing machine $M$, $L(M)$ denotes the set of strings accepted
by $M$, and the notation $M(x)$ means ``$M$ on input~$x$.''  For any
oracle Turing machine $M$ and any oracle set $A$, $L(M^A)$ denotes the
set of strings accepted by $M$ relative to $A$, and the notation
$M^A(x)$ means ``$M^A$ on input~$x$.''  Without loss of generality, we
assume each NPM and NPOM (in our standard enumeration of such
machines) $M$ has the property that for every~$n$, there is an 
integer~$\ell_n$ such that, for every~$x$ of length~$n$, every path of $M(x)$
is of length~$\ell_n$, and furthermore, in the case of oracle
machines, that~$\ell_n$ is independent of the oracle.  Let $A$ and $B$
be sets.  We say $A$ is {\em Turing reducible\/} to $B$ (denoted by
$A\leq_{T}^{p} B$ or $A\in \p^B$) if there is a DPOM $M$ such that
$A=L(M^B)$.  A set $B$ is {\em Turing-hard\/} for a complexity class
${\cal C}$ if for all $A\in {\cal C}$, $A\leq_{T}^{p} B$.  A set $B$ is 
{\em Turing-complete\/} for~$\tweak{\cal C}$ if $B$ is Turing-hard 
for $\tweak{\cal C}$ and $B \in {\cal C}$.

\section{Boolean Hierarchies over Classes Closed Under {Intersection}} 
\label{section:BH}

The Boolean hierarchy is a natural extension of the classes
NP~\cite{coo:c:theorem-proving,lev:j:universal} and \mbox{$\mbox{DP}
  \equalsdef \np \inter \conp$}~\cite{pap-yan:j:dp}. Both NP and DP
contain natural problems, as do the levels of the Boolean hierarchy.
For example, graph minimal uncolorability is known to be complete
for~DP~\cite{cai-mey:j:dp}. Note that DP clearly is closed under
intersection, but is not closed under union unless the polynomial
hierarchy collapses (due to~\cite{kad:joutdatedbychangkadin:bh}, see
also~\cite{cha-kad:t:characteristic,cha:thesis}).

\begin{definition}\label{def-bh}
\cite{cai-gun-har-hem-sew-wag-wec:j:bh1,koe-sch-wag:j:diff,hau:b:sets}\quad
Let $\tweak{\cal K}$ be any class of sets.
\begin{enumerate}
\item The {\em Boolean (``alternating sums'') hierarchy over} ${\cal K}$:
$$\c{1}({\cal K})\equalsdef{\cal K},\ \ 
\c{k}({\cal K})\equalsdef\left\{ \begin{array}{ll}
  \c{k-1}({\cal K})\union {\cal K}      & \mbox{if $k$ odd}\\
  \c{k-1}({\cal K})\inter \co{\cal K} & \mbox{if $k$ even}
\end{array}
\right.
,\ k\geq 2,\ \ 
\ch({\cal K})\equalsdef\bigcup_{k\geq 1}\c{k}({\cal K}).$$

\item The {\em nested difference hierarchy over} ${\cal K}$:
$$\d{1}({\cal K})\equalsdef{\cal K},\ \ 
  \d{k}({\cal K})\equalsdef{\cal K} \minus 
                                        \d{k-1}({\cal K}),\ k\geq 2,\ \ 
  \newdh({\cal K})\equalsdef\bigcup_{k\geq 1}\d{k}({\cal K}).$$

\item 
The {\em Hausdorff (``union of differences'') hierarchy over} 
${\cal K}$:\footnote{%
\protect\singlespacing
Hausdorff hierarchies~(\protect\cite{hau:b:sets}, see
\protect\cite{cai-gun-har-hem-sew-wag-wec:j:bh1,ber-bru-jos-sit-you:c:gen,gun-nas-wec:c:counting-survey},
respectively, for applications to NP, R, and
$\mbox{C}\!\!\!\!=\!\!\!\mbox{P}$) are interesting both in the case
where, as in the definition here, the sets are arbitrary sets from
$\tweak\cal K$, and, as is sometimes used in definitions, the sets from
$\tweak\cal K$ are required to satisfy additional containment conditions.
For classes closed under union and intersection, such as NP, the two
definitions are identical, level by level (\protect\cite{hau:b:sets},
see also~\protect\cite{cai-gun-har-hem-sew-wag-wec:j:bh1}).  In this
paper, as, e.g., UP, is not known to be closed under union, the
distinction is nontrivial.
}
$$\e{1}({\cal K}) \equalsdef {\cal K},\ \ 
\e{2}({\cal K}) \equalsdef {\cal K} \minus {\cal K},\ \ 
\e{k}({\cal K}) \equalsdef \e{2}({\cal K}) \union 
                                        \e{k-2}({\cal K}),\ \ k > 2,\ \
\eh({\cal K}) \equalsdef \bigcup_{k\geq 1}\e{k}({\cal K}).$$

\item The {\em symmetric difference hierarchy over} ${\cal K}$:
$$\sd{1}({\cal K})\equalsdef{\cal K},\ \ 
  \sd{k}({\cal K})\equalsdef\sd{k-1}({\cal K})\XOR {\cal K},\ k\geq 2,\ \ 
  \sdh({\cal K})\equalsdef\bigcup_{k\geq 1}\sd{k}({\cal K}).$$
\end{enumerate}
\end{definition}

It is easily seen that for any $\mbox{X}$ chosen from $\{\mbox{C, D,
  E, SD}\}$, if $\tweak{\cal K}$ contains $\emptyset$ and $\Sigma^*$,
then for any $k\geq 1$,
$$\mbox{X}_{k}({\cal K}) \cup \co\mbox{X}_{k}({\cal K}) \seq
\mbox{X}_{k+1}({\cal K}) \cap \co\mbox{X}_{k+1}({\cal K}).$$

The following fact is shown by an easy induction on $n$.

\begin{fact}\label{ch(coup)=dh(up)}
\quad
For every class ${\cal K}$ of sets and every $n\geq 1$, 
(a)~$\d{2n-1}({\cal K})=\co\c{2n-1}(\co{\cal K})$, and
(b)~$\d{2n}({\cal K})=\c{2n}(\co{\cal K})$.
\end{fact}

\noindent
{\bf Proof.} \quad
The base case holds by definition.
Suppose (a) and (b) to be true for $n\geq 1$. Then,
\[
\begin{array}{lclcl}
\d{2n+1}({\cal K})  
   & = & {\cal K}\inter (\co{\cal K}\union \d{2n-1}({\cal K})) 
   & \stackrel{\mbox{\protect\scriptsize hyp.}}{=} & 
         {\cal K}\inter (\co{\cal K}\union \co\c{2n-1}(\co{\cal K}))\\
   & = & {\cal K}\inter \co({\cal K}\inter \c{2n-1}(\co{\cal K}))
   & = & {\cal K}\inter \co\c{2n}(\co{\cal K})\\
   & = & \co(\co{\cal K}\union \c{2n}(\co{\cal K}))
   & = & \co\c{2n+1}(\co{\cal K})
\end{array}
\]
shows (a) for $n+1$, and
\[
\begin{array}{lclclcl}
\d{2n+2}({\cal K}) 
   & = & {\cal K} \minus ({\cal K} \minus \d{2n}({\cal K}))
   & \stackrel{\mbox{\protect\scriptsize hyp.}}{=} & 
         {\cal K}\inter (\co{\cal K}\union \c{2n}(\co{\cal K}))
   & = & \c{2n+2}(\co{\cal K})
\end{array}
\]
shows (b) for $n+1$.~\hfill$\Box$

\begin{corollary}~
$\ch(\up) = \co\ch(\up) = \newdh(\coup)$ and 
$\ch(\coup) = \co\ch(\coup) = \newdh(\up)$.
\end{corollary}

We are interested in the Boolean hierarchies over classes closed under
intersection (but perhaps not under union or complementation), such as
UP, US, and DP\@.  We state our theorems in terms of the class of
primary interest to us in this paper, UP\@.  However, many apply to
any nontrivial class (i.e., any class containing ${\Sigma}^*$ and
$\emptyset$) closed under intersection (see
Theorem~\ref{t:general-case}).  Although it has been proven in
\cite{cai-gun-har-hem-sew-wag-wec:j:bh1} and \cite{koe-sch-wag:j:diff}
that all the standard normal forms of Definition~\ref{def-bh} coincide
for \np,\footnote{%
\protect\singlespacing
Due essentially to its closure 
under union and intersection, and this reflects 
a more general behavior of classes closed under
union and intersection, as studied 
by Bertoni et al.~(\protect\cite{ber-bru-jos-sit-you:c:gen},
see 
also~\protect\cite{hau:b:sets,cai-gun-har-hem-sew-wag-wec:j:bh1,koe-sch-wag:j:diff,cha-kad:t:characteristic,cha:thesis}).}
the situation for \up\ 
seems to be different, as \up\ is probably not closed under union. 
(The closure of \up\ under intersection is straightforward.)
Thus, all the relations among those normal forms have to be
reconsidered for UP\@. 

We first prove that the symmetric difference hierarchy over UP (or any
class closed under intersection) equals the Boolean closure.  Though
K\"obler, Sch\"oning, and Wagner~\cite{koe-sch-wag:j:diff}
proved this for NP, their proof
gateways through a class whose proof of equivalence to the Boolean closure 
uses closure under union, and thus the following result is not implicit
in their paper.

\begin{thm}\label{sdh=bc}
\quad
$\sdh(\up) = \bcup$.
\end{thm}

\noindent
{\bf Proof.} \quad
The inclusion from left to right is clear.
For the converse inclusion, it is sufficient to show that
$\sdh(\up)$ is closed under all Boolean operations, as
$\bcup$, by definition, is the smallest class of sets that
contains \up\ and is closed under all Boolean operations.
Let $L$ and $L^{'}$ be arbitrary sets in $\sdh(\up)$. Then,
for some $k, \ell \geq 1$, there are sets 
$A_1,\ldots ,A_k,B_1,\ldots ,B_{\ell}$ in \up\ representing $L$ and $L^{'}$:
$$L=A_1\sym \cdots \sym A_k \mbox{\ \ and\ \ }
  L^{'}=B_1\sym \cdots \sym B_{\ell}.$$
So
\[
L\cap L^{'} = \left({\Large \sym}_{i=1}^{k}A_i\right)\cap
                \left({\Large \sym}_{j=1}^{\ell}B_j\right)
            = {\Large \sym}_{i\in\{1,\ldots ,k\},\,
                j\in\{1,\ldots ,\ell\}}(A_i\cap B_j),               
\]
and since \up\ is closed under intersection and $\sdh(\up)$ is
(trivially) closed under symmetric difference, we clearly have that
$L\cap L^{'}\in \sdh(\up)$.
Furthermore, since $\overline{L}=\Sigma^*\sym L$ implies that 
$\overline{L}\in \sdh(\up)$, $\sdh(\up)$ is closed under
complementation. Since all Boolean operations can be 
represented in terms of 
complementation and
intersection, our proof is complete.~\hfill$\Box$

\medskip

Next, we show that for any class closed under intersection,
instantiated below to the case of UP, the Boolean (alternating sums)
hierarchy over the class equals the Boolean closure of the class.
Our proof is inspired by the techniques used to prove equality in 
the case where closure under union may be assumed.

\begin{thm}\label{t:ch-bc}
\quad
$\ch(\up) = \bcup$.
\end{thm}

\noindent
{\bf Proof.} 
\quad
We will prove that $\sdh(\up) \subseteq \ch(\up)$.
By Theorem~\ref{sdh=bc}, this will suffice.  

Let $L$ be any set in $\sdh(\up)$.  Then there is a $k>1$ (the case
$k=1$ is trivial) such that $L\in \sd{k}(\up)$. Let $U_{1},\ldots ,U_{k}$
be the witnessing \up\ sets; that is, 
$L=U_{1}\sym U_{2}\sym \cdots \sym U_{k}$.
By 
the inclusion-exclusion rule,
$L$ satisfies the equalities below.
For odd $k$,
\begin{eqnarray*}
L & = & \left( \cdots \left( \left(
        (U_{1}\cup U_{2}\cup \cdots \cup U_{k}) \cap
        \left( \overline{\bigcup_{j_{1}<j_{2}} \left( 
        U_{j_{1}}\cap U_{j_{2}}
        \right)} \right) \right) \right. \right. \cup \\
  &   & \left. \left. \left( \bigcup_{j_{1}<j_{2}<j_{3}}\left( 
        U_{j_{1}}\cap U_{j_{2}}\cap U_{j_{3}}
        \right) \right) \right) \cap \cdots  \cup
        \left( \bigcup_{j_{1}< \cdots <j_{k}}\left( 
        U_{j_{1}}\cap \cdots \cap U_{j_{k}}
        \right) \right) \right) ,
\end{eqnarray*}
where each subscripted $j$ term must belong
to $\{1,\ldots ,k\}$.
For even $k$, we similarly have:
\begin{eqnarray*}
L & = & \left( \cdots \left(\left(
        (U_{1}\cup U_{2}\cup \cdots \cup U_{k}) \cap
        \left( \overline{\bigcup_{j_{1}<j_{2}} \left( 
        U_{j_{1}}\cap U_{j_{2}}
        \right)} \right) \right) \right. \right. \cup \\
  &   & \left. \left. \left( \bigcup_{j_{1}<j_{2}<j_{3}}\left( 
        U_{j_{1}}\cap U_{j_{2}}\cap U_{j_{3}}
        \right) \right) \right) \cap \cdots  \cap
        \left( \overline{\bigcup_{j_{1}< \cdots <j_{k}}\left( 
        U_{j_{1}}\cap \cdots \cap U_{j_{k}}
        \right) }\right) \right)  .
\end{eqnarray*}
For notational convenience, let us use $A_1,\ldots ,A_k$ to represent
the respective terms in the above expressions (ignoring the
complementations).  By the closure of UP under intersection, each
$A_i$, $1\leq i\leq k$, is the union of ${k \choose i}$ UP sets
$B_{i,1}$, $\ldots$, $B_{i,{k \choose i}}$.  Using the fact that
$\emptyset$ is clearly in UP, we can easily turn the union of $n$
arbitrary UP sets (or the intersection of $n$ arbitrary coUP sets)
into an alternating sum of $2n-1$ UP sets. So for instance, $A_1 =
U_{1}\cup U_{2}\cup \cdots \cup U_{k}$ can be written
$$\left( \cdots \left( \left( \left( U_{1} \cap
\overline{\emptyset}\right) \cup U_{2}\right) \cap
\overline{\emptyset}\right) \cup \cdots \cup U_{k}\right),$$ 
call this $C_1$. Clearly, $C_1 \in \c{2k-1}(\up)$.  To transform the
above representation of $L$ into an alternating sum of UP sets, we
need two (trivial) transformations holding for any $m\geq 1$ and for
arbitrary sets $S$ and $T_1,\ldots ,T_m$:
\begin{eqnarray}
S\cap \left( \overline{T_1\cup T_2\cup \cdots \cup T_m} \right)
& = &
\left( \cdots \left(\left(S\cap \overline{T_1}\right)\cap 
\overline{T_2}\right) \cap \cdots \right) \cap \overline{T_m} 
\label{equ:trivial-1} \\
S\cup \left( T_1\cup T_2\cup \cdots \cup T_m \right)
& = &
\left( \cdots \left( \left(S\cup T_1 \right) \cup T_2 
\right) \cup \cdots \right) \cup T_m.  \label{equ:trivial-2} 
\end{eqnarray}
\noindent Using (\ref{equ:trivial-1}) with $S=C_1$ and
$T_1=B_{2,1},\ldots ,T_m=B_{2,{k \choose 2}}$ and the fact that
$\emptyset$ is in UP, $A_1\cap \overline{A_2}$ can be transformed into
an alternating sum of UP sets, call this $C_2$.  Now apply
(\ref{equ:trivial-2}) with $S=C_2$ and $T_1=B_{3,1},\ldots
,T_m=B_{3,{k \choose 3}}$ to obtain, again using that $\emptyset$ is
in \up\@, an alternating sum $C_3 = \left(A_1\cap
\overline{A_2}\right) \cup A_3$ of UP sets, and so on. Eventually,
this procedure of alternately applying (\ref{equ:trivial-1}) and
(\ref{equ:trivial-2}) will yield an alternating sum $C_{k}$ of sets in
UP that equals $L$. Thus, $L\in \ch(\up)$.~\hfill$\Box$

\begin{corollary} \label{cor:sxNEW}
\quad
$\sdh(\up)$ and $\ch(\up)$ are both closed under all Boolean operations.
\end{corollary}

Note that the proofs of Theorems~\ref{t:ch-bc} and~\ref{sdh=bc} 
implicitly give a recurrence yielding an
upper bound on the level-wise containments.
We find the issue
of equality to $\bcup$, or lack thereof, to be the central issue, and
thus we focus on that. Nonetheless, we point out in the corollary below
that losing the 
assumption of closure under union seems to have exacted a price:
though the hierarchies $\sdh({\up})$ and $\ch({\up})$ are indeed
equal, the above proof embeds $\sd{k}(\up)$ in an exponentially
higher level of the C~hierarchy. Similarly, the proof of
Theorem~\ref{sdh=bc} embeds $\c{k}(\up)$ in an exponentially higher
level of $\sdh(\up)$\@.

\begin{cor}{\bf (to the proofs of Theorems~\ref{t:ch-bc} and~\ref{sdh=bc})}~
\label{cor:level-wise}
\begin{enumerate}
\item For each $k \geq 1$, $\sd{k}(\up) \seq \c{2^{k+1}-k-2}(\up)$.

\item For each $k \geq 1$, $\c{k}(\up) \seq \sd{T(k)}(\up)$, where
$
T(k) = \left\{ \begin{array}{ll}
       2^{k} - 1       & \mbox{if $k$ is odd}\\
       2^{k} - 2       & \mbox{if $k$ is even.}
\end{array}
\right.
$
\end{enumerate}
\end{cor}

\noindent
{\bf Proof.} 
\quad
For an $\sd{k}(\up)$ set $L$ to be placed into the $R(k)$th level of
$\ch(\up)$, $L$ is represented (in the proof of Theorem~\ref{t:ch-bc})
as an alternating sum of $k$ terms $A_1,\ldots , A_k$, each $A_i$
consisting of ${k \choose i}$ UP sets $B_{i,j}$. In the subsequent
transformation of $L$ according to the equations (\ref{equ:trivial-1})
and (\ref{equ:trivial-2}), each $A_i$ requires as many as ${k \choose i} 
- 1$ additional terms $\emptyset$ or $\overline{\emptyset}$,
respectively, to be inserted, and each such insertion brings us one
level higher in the C~hierarchy. Thus,
\[
R(k) = \sum_{i=1}^{k} {k \choose i} + \left( {k \choose i} -1 \right)
     = - k + 2 \sum_{i=1}^{k} {k \choose i} 
     = 2^{k+1}-k-2.
\]
A close inspection of the proof of $\c{k}(\up) \seq \sd{T(k)}(\up)$
according to Theorem~\ref{sdh=bc} leads to the recurrence:
\[
\begin{array}{lcl}
T(1) = 1 & \mbox{ and } & 
T(k) = \left\{ \begin{array}{ll}
        2T(k-1) + 3 & \mbox{if $k > 1$ is odd}\\
        2T(k-1)     & \mbox{if $k > 1$ is even,}
\end{array}
\right.
\end{array}
\]
since any set $L \in \c{k}(\up)$ can be represented by sets $A \in
\c{k-1}(\up)$ and $B \in \up$ as follows:
\[
\begin{array}{llllllll}
L & = & A \cup B & = & \overline{\overline{A} \cap \overline{B}} & = &
\Sigma^* \xor \left( \left( \Sigma^* \xor A \right) \cap \left( 
\Sigma^* \xor B \right) \right) & \hspace*{.5cm}\mbox{if $k$ is odd,}\\
L & = & A \cap \overline{B} & = & A \cap \left( 
\Sigma^* \xor B \right)  &  &   & \hspace*{.5cm}\mbox{if $k$ is even.}
\end{array}
\]
The above recurrence is in (almost) closed form:
\begin{eqnarray*}
T(k) & = & \left\{ \begin{array}{ll}
       2^{k} - 1       & \mbox{if $k \geq 1$ is odd}\\
       2^{k} - 2       & \mbox{if $k \geq 1$ is even,}
\end{array}
\right.
\end{eqnarray*}
as can be proven by induction on $k$ (we omit the trivial induction
base): For odd $k$ (i.e., $k = 2n-1$ for $n \geq 1$), assume $T(2n-1)
= 2^{2n-1} - 1$ to be true. Then,
\[
T(2n+1) = 2T(2n) + 3 = 4T(2n-1) + 3 \stackrel{\mbox{\protect\scriptsize hyp.}}{=} 
4\left( 2^{2n-1} - 1\right) + 3 = 2^{2n+1} - 1.
\]
For even $k$ (i.e., $k = 2n$ for $n \geq 1$),
assume $T(2n) = 2^{2n} - 2$ to be true. Then,
\[
T(2n+2) = 2T(2n+1) = 2(2T(2n) + 3) \stackrel{\mbox{\protect\scriptsize hyp.}}{=} 
4\left( 2^{2n} - 2\right) + 6 = 2^{2n+2} - 2.
\ \ \Box
\]

\begin{remark}\label{rem:level-wise}
  \quad The upper bound in the second part of the above
  proof can be slightly improved using the fact that $\Sigma^* \xor
  \Sigma^* \xor A = \emptyset \xor A = A$ for any set $A$. This gives
  the recurrence:
\[
\begin{array}{lcl}
T(1) = 1 & \mbox{ and } & 
T(k) = \left\{ \begin{array}{ll}
        2T(k-1) + 1 & \mbox{if $k > 1$ is odd}\\
        2T(k-1)     & \mbox{if $k > 1$ is even,}
\end{array}
\right.
\end{array}
\]
or, equivalently, $T(1) = 1$, $T(2) = 2$, and $T(k) = 2^{k-1} +
T(k-2)$ for $k \geq 3$. Though this shows that the upper bound given in the
above proof is not optimal, the new bound is not a strong improvement, 
as it still embeds $\c{k}(\up)$ in an exponentially higher
level of $\sdh(\up)$\@. We propose as an interesting task the establishment 
of {\em tight\/} level-wise containments, at
least up to the limits of relativizing techniques, between the hierarchies
$\sdh(\up)$ and $\ch(\up)$, both of which capture the Boolean closure of UP.

We conjecture that there is some
relativized world in which an exponential increase (though less dramatic
than the particular exponential increase of
Corollary~\ref{cor:level-wise}) indeed is necessary.
\end{remark}

Theorem~\ref{dinc} below shows that each level of the nested
difference hierarchy is contained in the same level of both the C and
the E hierarchy.  Surprisingly, it turns out (see
Theorem~\ref{t:oracle} below) that, relative to a recursive oracle,
even the fourth level of $\ch(\up)$ and the third level of $\sdh(\up)$
are not subsumed by any level of the $\eh(\up)$ hierarchy.
Consequently, neither the D nor the E normal forms of
Definition~\ref{def-bh} capture the Boolean closure of UP\@.

\begin{thm}\label{dinc}
\quad
For every $k\geq 1$, $\d{k}(\up) \seq \c{k}(\up)\cap \e{k}(\up)$.
\end{thm}

\noindent
{\bf Proof.} \quad
For the first inclusion, by 
\cite[Proposition~2.1.2]{cai-hem:t:boolean}, each set 
$L\in \d{k}(\up)$ can be represented as
$$L=A_1-(A_2-(\cdots(A_{k-1} -A_k)\cdots )),$$
where $A_i=\bigcap_{1\leq j\leq i}L_j$, $1\leq i\leq k$, and
the $L_j$'s are the original \up\ sets representing $L$.
Note that since the proof of 
\cite[Proposition~2.1.2]{cai-hem:t:boolean} only uses intersection,
the sets $A_i$ are in \up\@.  A 
special case of \cite[Proposition~2.1.3]{cai-hem:t:boolean}
says that sets in 
$\d{k}({\rm UP})$ 
via decreasing chains such as the $A_i$ are in
$\c{k}({\rm UP})$, and so 
$L\in \c{k}(\up)$.

The proof of the second inclusion is done by induction on the odd 
and even levels separately. 
The induction base follows by definition in either case.
For odd levels, assume $\d{2n-1}(\up)\seq \e{2n-1}(\up)$ to be valid, 
and let $L$ be any set in 
$\d{2n+1}(\up)=\up \minus (\up \minus \d{2n-1}(\up))$.
By our inductive hypothesis, $L$ can be represented as
$$L=A-\left( B-\left( \bigcup_{i=1}^{n-1}\left( C_i\cap 
  \overline{D_i}\right) \cup E\right) \right) ,$$
where $A,B,C_i,D_i$, and $E$ are sets in \up. Thus,
\begin{eqnarray*}
L & = & A\cap\left(\overline{B\cap\left(\overline{\bigcup_{i=1}^{n-1}
        \left(C_i\cap \overline{D_i}\right)\cup E}\right)}\right)\\
  & = & A\cap\left(\overline{B}\cup\left(\bigcup_{i=1}^{n-1}
        \left(C_i\cap \overline{D_i}\right)\cup E\right)\right)\\
  & = & (A\cap \overline{B})\cup \left(\bigcup_{i=1}^{n-1}
        A\cap C_i\cap \overline{D_i}\right)\cup (A\cap E)\\
  & = & \left(\bigcup_{i=1}^{n}F_i\cap \overline{D_i}\right)\cup G,
\end{eqnarray*}
where $F_i=A\cap C_i$, for $1\leq i\leq n-1$, $F_n=A$, $D_n=B$, and
$G=A\cap E$. Since \up\ is closed under intersection, each of these
sets is in \up. Thus, $L\in \e{2n+1}(\up)$.  The proof for the even
levels is analogous except that the set $E$ is dropped.~\hfill$\Box$

\medskip

Note that most of the above proofs used only the facts that the class
is closed under intersection and contains ${\Sigma}^*$ and
$\emptyset$:

\begin{theorem} 
\label{t:general-case}
\quad
Theorems~\ref{sdh=bc},~\ref{t:ch-bc}, and~\ref{dinc} and 
Corollaries~\ref{cor:sxNEW} and~\ref{cor:level-wise} apply to all
classes that contain ${\Sigma}^*$ and $\emptyset$ and are closed under
intersection.
\end{theorem}

\begin{remark} \quad
  Although DP is closed under intersection but seems to lack closure
  under union (unless the polynomial hierarchy collapses to
  DP~\cite{kad:joutdatedbychangkadin:bh,cha-kad:t:characteristic,cha:thesis}) 
  and thus
  Theorem~\ref{t:general-case} in particular applies to DP, we note
  that the known results about Boolean hierarchies over
  NP~\cite{cai-gun-har-hem-sew-wag-wec:j:bh1,koe-sch-wag:j:diff} in
  fact even for the DP case imply stronger results than those given by
  our Theorem~\ref{t:general-case}, due to the very special 
  structure of DP\@. Indeed, since, e.g., $\e{k}(\dnp)
  = \e{2k}(\np)$ for any $k \geq 1$ (and the same holds for the other
  hierarchies), it follows immediately that all the level-wise
  equivalences among the Boolean hierarchies (and also their ability to
  capture the Boolean closure) that are known to hold for NP also hold
  for DP even in the absence of the assumption of closure under union.
  This appears to contrast with the UP case (see
  Remark~\ref{rem:level-wise}).
\end{remark}

The following combinatorial lemma will be useful in proving
Theorem~\ref{t:oracle}.

\begin{lem}\label{no-party}
\cite{cai-hem-vys:b:promise}
\quad
Let $G = (S, T, E)$ be any directed bipartite graph
with out-degree bounded by $d$ for all vertices.
Let $S' \subseteq S$ and $T' \subseteq T$
be subsets such that \linebreak[4]
$S' \supseteq \{s \in S\,|\,(\exists t\in T) \,
[\<s,t\>\in E]\}$,
and 
$T' \supseteq \{t \in T\,|\,
(\exists s  \in S)\, [  \<t,s\>\in E]\}$.
Then either:
 \vspace{-.1in}
 \begin{enumerate}
 \item
 $\|S'\|\leq 2d$, or
  \item
 $\|T'\|\leq 2d$, or
 \item \label{no-party-part-3}
 $(\exists s \in S')\,(\exists t \in T')\,[\<s,t\>
\not \in E \,\wedge\,
   \<t,s\>\not \in E]$.
 \end{enumerate}
 \end{lem}

 For papers concerned with oracles separating internal levels of
 Boolean hierarchies over classes other than those of this paper, we
 refer the reader
 to~(\cite{cai-gun-har-hem-sew-wag-wec:j:bh1,cai:c:boolean-1,gun-nas-wec:c:counting-survey,bru-jos-you:j:strong,cro:t:boolean},
 see also~\cite{gun-wec:j:count}).  Theorem~\ref{t:oracle} is optimal,
 as clearly $\c{3}(\up)\seq\eh(\up)$ and $\sd{2}(\up)\seq\eh(\up)$,
 and both these containments relativize.

\begin{thm}\label{t:oracle}
\quad
There are recursive oracles $A$ and 
$D$ (though we may take $A = D$) such that
\begin{enumerate} 
\item \label{oracle-1} $\c{4}(\up^{A})\not\seq \mbox{EH}(\up^{A})$, and
\item \label{oracle-2} $\sd{3}(\up^{D})\not\seq \mbox{EH}(\up^{D}).$
\end{enumerate}
\end{thm}

\begin{corollary}
\quad
There is a recursive oracle $A$ such that
\begin{enumerate}
\item
$\mbox{EH}(\up^A) \neq \mbox{BC}(\up^A)$ and
$\mbox{DH}(\up^A) \neq \mbox{BC}(\up^A)$,\footnote{%
\protect\singlespacing
As Fact~\ref{ch(coup)=dh(up)} shows that
$\mbox{DH}(\up)=\mbox{CH}(\co\up)$, this oracle $A$ also separates the
Boolean (alternating sums) hierarchy over coUP from the fourth level
of the same hierarchy over UP and, thus, from $\bcup$.
}
and

\item $\mbox{EH}(\up^A)$ and $\mbox{DH}(\up^A)$ are not closed under
  all Boolean operations.
\end{enumerate}
\end{corollary}

\medskip

\noindent {\bf Proof of Theorem~\ref{t:oracle}.} 
\quad
Although the theorem claims there is an 
oracle keeping 
$\c{4}(\up)$ from 
being contained in any level of $\eh(\up)$,
we will only prove that for any fixed $k$ we can 
ensure that $\c{4}(\up)$ is not contained in $\e{k}(\up)$,
relative to some oracle $A^{(k)}$.
In the standard way, by interleaving diagonalizations, the sequence
of oracles, $A^{(k)}$, can be combined into a single 
oracle,~$A$, that fulfills the claim of the theorem. 
An analogous comment holds for the second claim of the theorem,
with a sequence of oracles
$D^{(k)}$ yielding a single oracle $D$.  Similarly, 
both statements of the theorem can be 
satisfied simultaneously via just 
one oracle, via interleaving with each other the constructions of $A$
and $D$.  Though below we construct just $A^{(k)}$ and $D^{(k)}$, 
as a notational shorthand we'll use $A$ and $D$ below to represent 
$A^{(k)}$ and $D^{(k)}$.

Before the actual construction of the oracles, we state some
preliminaries that apply to the proofs of both statements in the
theorem.

For any $n\geq 0$ and any string $v\in \Sigma^{\leq n}$, define
$S_{v}^{n} \equalsdef \{vw\,|\,vw\in \Sigma^n\}$. The
sets $S_{v}^{n}$ are used to distinguish between different segments of
$\Sigma^n$ in the definition of the test languages, $L_A$ and $L_D$.

Fix any standard enumeration of all NPOMs.  Fix any $k > 0$. We need
only consider even levels of $\eh(\up)$, as each odd level is contained
in some even level.  Call any collection of $2k$ NPOMs, $H =
\<N_{1,1},\ldots , N_{k,1}, N_{1,2} ,\ldots , N_{k,2}\>$, a potential
(relativized) $\e{2k}(\up)$ machine, and for any oracle $X$, define
its language to be:
\[
L(H^X) \equalsdef \bigcup_{i=1}^{k}\left(L(N_{i,1}^{X})-L(N_{i,2}^{X})\right).
\] 
If for some fixed oracle $Y$, a potential (relativized) $\e{2k}(\up)$
machine $H^Y$ has the property that each of its underlying NPOMs with oracle
$Y$ is unambiguous, then $L(H^Y)$ indeed is in $\e{2k}(\up^Y)$\@.
Clearly, our enumeration of all NPOMs induces an enumeration of all
potential $\e{2k}(\up)$ oracle machines.  For $j \geq 1$, let $H_j$ be
the $j$th machine in this enumeration.  Let $p_j$ be a polynomial
bounding the length of the computation paths of each of $H_j$'s
underlying machines (and thus bounding the number of and length of the
strings they each query).  As a notational convenience, we
henceforward will use $H$ and $p$ as shorthands for $H_j$ and $p_j$,
and we will denote the underlying NPOMs by $N_{1,1},\ldots , N_{k,1},
N_{1,2} ,\ldots , N_{k,2}$.

The oracle $X$, where $X$ stands for $A$ or $D$, is constructed 
in stages, $X=\bigcup_{j\geq 1}X_j$.
In stage $j$, we diagonalize against $H$ by satisfying the following
requirement $R_j$ for every $j\geq 1$:
\begin{description}
\item[$R_j:$] Either there is an $n > 2$ and an $i$, $1 \leq i \leq k$, 
              such that one of 
              $N_{i,1}^{X_j}$ or $N_{i,2}^{X_j}$ on input $0^n$ is
              ambiguous (thus, $H$ is in fact not an $\e{2k}(\up)$ 
              machine relative to $X$), or $L(H^X)\neq L_X$.
\end{description}

Let $X_j$ be the set of strings contained in $X$ by the end of
stage $j$, and let $X^{'}_{j}$ be the set of strings forbidden 
membership in $X$ during stage $j$. The restraint function $r(j)$ 
will satisfy the condition that at no later stage will strings of length 
smaller than $r(j)$ be added to $X$\@.  Also, 
our construction will ensure that 
$r(j)$ is so large that $X_{j-1}$ contains no strings
of length greater than $r(j)$.
Initially, both $X_0$ and $X^{'}_{0}$ are empty, and $r(1)$ is set to be 2.

\smallskip

We now start the proof of Part \ref{oracle-1}
of the theorem.   Define the test language:
$$L_A \equalsdef
 \{0^n\,|\,(\exists x)\,[x\in S_{0}^{n}\cap A]\,\wedge\,
           (\forall y)\,[y\not\in S_{10}^{n}\cap A]\,\wedge\,
           (\forall z)\,[z\not\in S_{11}^{n}\cap A]\}.$$

Clearly, $L_A$ is in \mbox{$\np^A \inter \conp^A \inter \conp^A$}. 
However, if we ensure in the
construction that the invariant $\|S_{v}^{n}\cap A\|\leq 1$
is maintained
for $v\in \{0,10,11\}$ and every $n\geq 2$, 
then $L_A$ is even in $\up^A\inter\co\up^A\inter\co\up^A$,
and thus in $\c{4}(\up^{A})$.
We now describe stage $j>0$ of the oracle construction.

\begin{description}
\item[Stage {\boldmath $j$}:]
Choose $n>r(j)$ so large that $2^{n-2}>3p(n)$.
\begin{description}
\item[Case 1:] $0^n\in L(H^{A_{j-1}})$. Since $0^n\not\in L_A$, we have
               $L(H^A)\neq L_A$.
\item[Case 2:] $0^n\not\in L(H^{A_{j-1}})$. Choose some $x\in S_{0}^{n}$
               and set $B_j:=A_{j-1}\cup \{x\}$.
\begin{description}
\item[Case 2.1:] $0^n\not\in L(H^{B_j})$. Letting $A_j:=B_j$
                 implies $0^n\in L_A$, so $L(H^A)\neq L_A$.
\item[Case 2.2:] $0^n\in L(H^{B_j})$. Then there is an $i$, 
                 $1\leq i\leq k$, such that $0^n\in L(N_{i,1}^{B_j})$
                 and $0^n\not\in L(N_{i,2}^{B_j})$. ``Freeze'' an
                 accepting path of $N_{i,1}^{B_j}(0^n)$ into 
                 $A^{'}_{j}$; that is, add those strings queried
                 negatively on that path to $A^{'}_{j}$, thus forbidding 
                 them from $A$ for all later stages. Clearly,
                 at most $p(n)$ strings are ``frozen.''
\begin{description}
\item[Case 2.2.1:] $\left(\exists z\in 
                   (S_{10}^n\cup S_{11}^n)-A^{'}_{j}\right)\,
                   \left[0^n\not\in L(N_{i,2}^{B_j\cup \{z\}})\right]$. \\
                   Choose any such $z$. Set $A_j:=B_j\cup \{z\}$.
                   We have $0^n\in L(H^A)-L_A$.
\item[Case 2.2.2:] $\left(\forall z\in 
                   (S_{10}^n\cup S_{11}^n)-A^{'}_{j}\right)\,
                   \left[0^n\in L(N_{i,2}^{B_j\cup \{z\}})\right]$. \\
                   To apply Lemma~\ref{no-party}, define a
                   directed bipartite graph $G=(S,T,E)$ by
                   $S \equalsdef S_{10}^n-A^{'}_{j}$, 
                   $T \equalsdef S_{11}^n-A^{'}_{j}$,  
                   and for each $s\in S$ and $t\in T$, $\<s,t\>\in E$
                   if and only if $N_{i,2}^{B_j\cup \{s\}}$ queries 
                   $t$ along its lexicographically first accepting
                   path, and $\<t,s\>\in E$ is defined analogously.
                   The out-degree of all vertices of $G$ is bounded 
                   by $p(n)$. By our choice of $n$,
                   $\min\{\|S\|, \|T\|\} \geq 2^{n-2} - p(n) > 2p(n)$,
                   and thus
                   alternative~\ref{no-party-part-3}
                   of Lemma~\ref{no-party} applies. Hence, there 
                   exist strings $s\in S$ and $t\in T$ such that 
                   $N_{i,2}^{B_j\cup \{s\}}(0^n)$ accepts on some path $p_s$
                   on which $t$ is not queried, and
                   $N_{i,2}^{B_j\cup \{t\}}(0^n)$ accepts on some path $p_t$
                   on which $s$ is not queried.
                   Since $p_s$ ($p_t$) changes from reject to accept exactly
                   by adding $s$~($t$) to the oracle, $s$~($t$) must have 
                   been queried on $p_s$~($p_t$). We conclude that 
                   $p_s \neq p_t$, and thus
                   $N_{i,2}^{B_j\cup \{s,t\}}(0^n)$ has at least
                   two accepting paths. Set $A_j:=B_j\cup \{s,t\}$.
\end{description}
\end{description}
\end{description}
\end{description}
In each case, requirement $R_j$ is fulfilled. Let 
$r(j+1)$ be $\max\{n,w_j\}$, where $w_j$ is the length of 
the largest string queried through stage $j$.\\
{\bf End of stage {\boldmath $j$}.}

\smallskip

We now turn to the proof of Part \ref{oracle-2} of the
theorem.  The test language here, $L_D$, is defined by:
$$L_D \equalsdef \left \{ 0^n\,
\begin{tabular*}{11.1cm}{|l}  %
$((\exists x)\,[x\in S_{0}^{n}\cap D]\,\wedge\,
  (\exists y)\,[y\in S_{10}^{n}\cap D]\,\wedge\,
  (\exists z)\,[z\in S_{11}^{n}\cap D])\,\vee $\\
$((\forall x)\,[x\not\in S_{0}^{n}\cap D]\,\wedge\,
  (\forall y)\,[y\not\in S_{10}^{n}\cap D]\,\wedge\,
  (\exists z)\,[z\in S_{11}^{n}\cap D])\,\vee $\\
$((\exists x)\,[x\in S_{0}^{n}\cap D]\,\wedge\,
  (\forall y)\,[y\not\in S_{10}^{n}\cap D]\,\wedge\,
  (\forall z)\,[z\not\in S_{11}^{n}\cap D])\,\vee $\\
$((\forall x)\,[x\not\in S_{0}^{n}\cap D]\,\wedge\,
  (\exists y)\,[y\in S_{10}^{n}\cap D]\,\wedge\,
  (\forall z)\,[z\not\in S_{11}^{n}\cap D])$
\end{tabular*}
\right\}.$$
Again, provided that the invariant $\|S_{v}^{n}\cap D\|\leq 1$ is 
maintained for $v\in \{0,10,11\}$ and every $n\geq 2$ 
throughout the construction, $L_D$ is clearly 
in $\sd{3}(\up^{D})$,
as for all sets $A$, $B$, and $C$,
$$A \sym B \sym C=
  (A\cap B\cap C)\cup (\overline{A}\cap \overline{B}\cap C)\cup 
  (A\cap \overline{B}\cap \overline{C})\cup
  (\overline{A}\cap B\cap \overline{C}).$$
Stage $j>0$ of the construction of $D$ is as follows.

\begin{description}
\item[Stage {\boldmath $j$}:]
Choose $n>r(j)$ so large that $2^{n-2}>3p(n)$.
\begin{description}
\item[Case 1:] $0^n\in L(H^{D_{j-1}})$. Since $0^n\not\in L_D$, we have
               $L(H^D)\neq L_D$.
\item[Case 2:] $0^n\not\in L(H^{D_{j-1}})$. Choose some $x\in S_{0}^{n}$
               and set $E_j:=D_{j-1}\cup \{x\}$.
\begin{description}
\item[Case 2.1:] $0^n\not\in L(H^{E_j})$. Letting $D_j:=E_j$
                 implies $0^n\in L_D$, so $L(H^D)\neq L_D$.
\item[Case 2.2:] $0^n\in L(H^{E_j})$. Then, there is an $i$, 
                 $1\leq i\leq k$, such that $0^n\in L(N_{i,1}^{E_j})$
                 and $0^n\not\in L(N_{i,2}^{E_j})$. ``Freeze'' an
                 accepting path of $N_{i,1}^{E_j}(0^n)$ into 
                 $D^{'}_{j}$. Again, at most $p(n)$ strings 
                 are ``frozen.''
\begin{description}
\item[Case 2.2.1:] $\left(\exists w\in 
                   (S_{10}^n\cup S_{11}^n)-D^{'}_{j}\right)\,
                   \left[0^n\not\in L(N_{i,2}^{E_j\cup \{w\}})\right]$.\\
                   Choose any such $w$ and set $D_j:=E_j\cup \{w\}$.
                   We have $0^n\in L(H^D)-L_D$.
\item[Case 2.2.2:] $\left(\forall 
                   w\in (S_{10}^n\cup S_{11}^n)-D^{'}_{j}\right)\,
                   \left[0^n\in L(N_{i,2}^{E_j\cup \{w\}})\right]$.\\
                   As before, Lemma~\ref{no-party} yields
                   two strings 
                   $s\in S_{10}^n-D^{'}_{j}$ and 
                   $t\in S_{11}^n-D^{'}_{j}$ such that 
                   $N_{i,2}^{E_j\cup \{s,t\}}(0^n)$ is ambiguous. 
                   Set $D_j:=E_j\cup \{s,t\}$.
\end{description}
\end{description}
\end{description}
\end{description}
Again, $R_j$ is always fulfilled. Define $r(j+1)$ as before.\\
{\bf End of stage {\boldmath $j$}.}  
\hfill$\Box$

\medskip

Finally, we note that a slight modification of the above proof
establishes the analogous result (of Theorem~\ref{t:oracle}) for the
case of US~\cite{bla-gur:j:unique-sat} (which is denoted 1NP
in~\cite{gun-wec:j:count,cro:t:boolean}).

\section{Sparse Turing-complete and Turing-hard Sets 
  for UP}\label{section:Turing-Complete-Sets}

In this section, we show some consequences of the existence of sparse
Turing-complete and Turing-hard sets for \up\@. This question has been
carefully investigated for the class
\np~\cite{kar-lip:c:nonuniform,hop:c:recent,ko-sch:j:circuit-low,bal-boo-sch:j:sparse,lon-sel:j:sparse,sch:b:complexity-structure,kad:j:pnplog}\@.\footnote{%
\protect\singlespacing
For reductions less flexible than Turing
reductions (e.g., $\leq_m^p$, $\leq_{btt}^p$, etc.), this issue has been
studied even more intensely (see, e.g., the 
surveys~\protect\cite{you:j:sparse,hem-ogi-wat:c:sparse}).
} 
Kadin showed that if there is a sparse $\leq_{T}^{p}$-complete set in~NP, 
then the polynomial hierarchy collapses to 
$\p^{\protect\scriptsize \np[\log]}$~\cite{kad:j:pnplog}.  
Due to the promise nature of~UP (in
particular, \up\ probably lacks complete sets \cite{har-hem:j:up}),
Kadin's proof does not seem to apply here. But does the existence of a
sparse Turing-complete set in \up\ cause at least some collapse of the
unambiguous polynomial hierarchy
(which was introduced recently in~\cite{nie-ros:c:up-hierarchy})?\footnote{%
\protect\singlespacing
Note that it is not known whether such a collapse
implies a collapse of PH\@.
Note also that Toda's~\cite{tod:j:psel} result on whether P-selective
sets can be truth-table-hard for UP does not imply such a 
collapse, as truth-table reductions are less flexible than
Turing reductions.
}

Cai, Hemachandra,~and Vysko\v{c}~\cite{cai-hem-vys:b:promise} observe
that ordinary Turing access to \up, as formalized by $\p^{\protect\scriptsize
  \up}$, may be too restrictive a notion to capture adequately one's
intuition of Turing access to unambiguous computation, since in that
model the oracle machine has to be unambiguous on {\em every\/}
input---even those the base DPOM never asks (on any of {\em its\/}
inputs).  To relax that unnaturally strong uniformity requirement they
introduce the class denoted $\p^\smallprup$,
in which NP oracles
are
accessed in a {\em guardedly\/} unambiguous manner,
a natural notion of access to unambiguous computation---suggested 
in the rather analogous case of $\np \cap \conp$ by
Grollmann and 
Selman~\cite{gro-sel:j:complexity-measures}---%
in which {\em only computations actually executed need be
  unambiguous}.  Lange, Niedermeier, and
Rossmanith~\cite{lan-ros:c:up-circuit-and-hierarchy}\cite[p.~483]{nie-ros:c:up-hierarchy}
generalize this approach to build up an entire hierarchy of
unambiguous computations in which the oracle levels are guardedly
accessed (Definition~\ref{d:hierarchies},
Part~\ref{def-part:promise})---the {\em promise unambiguous polynomial
  hierarchy}.

\begin{definition}~\label{d:hierarchies}
\begin{enumerate}
\item
The {\em polynomial hierarchy}~\cite{mey-sto:b:reg-exp-needs-exp-space,sto:j:poly} is defined as follows:\\
$\sigp{0}\equalsdef\p$, \
$\delp{0}\equalsdef\p$, \
$\sigp{k}\equalsdef\np^{\sigp{k-1}}$, \
$\pip{k}\equalsdef\co\sigp{k}$,  \
$\delp{k}\equalsdef\p^{\sigp{k-1}}$, $k\geq 1$, \ and
$\ph \equalsdef\bigcup_{k\geq 0}\sigp{k}$.

\item The {\em unambiguous polynomial 
hierarchy}~\cite{nie-ros:c:up-hierarchy} is defined as follows:\\
$\u\sigp{0}\equalsdef\p$, 
$\u\delp{0}\equalsdef\p$,
$\u\sigp{k}\equalsdef\up^{\protect\scriptsize \u\sigp{k-1}}$, 
$\u\pip{k}\equalsdef\co\u\sigp{k}$,
$\u\delp{k}\equalsdef\p^{\protect\scriptsize \u\sigp{k-1}}$, $k\geq 1$, 
and $\u\ph \equalsdef\bigcup_{k\geq 0}\u\sigp{k}$.

\item \label{def-part:promise}
The {\em promise unambiguous polynomial 
hierarchy}~(\cite{lan-ros:c:up-circuit-and-hierarchy}\cite[p.~483]{nie-ros:c:up-hierarchy}) 
is defined as follows:
$\pru\sigp{0}\equalsdef\p$, $\pru\sigp{1}\equalsdef\up$, and 
for $k\geq 2$, $L\in \pru\sigp{k}$ if and only if 
$L\in \sigp{k}$ via NPOMs $N_1,\ldots ,N_k$ satisfying for all 
inputs $x$ and every $i$, $1\leq i\leq k-1$, that if $N_i$ asks some 
query $q$ during the computation of $N_1(x)$, 
then $N_{i+1}(q)$ with oracle $L(N_{i+2}^{L(N_{i+3}^{\.^{\.^{L(N_{k})}}})})$
has at most one accepting path.
$\pruph\equalsdef\bigcup_{k\geq 0}\pru\sigp{k}$.
The classes $\pru\delp{k}$ and $\pru\pip{k}$, $k\geq 0$, are defined
analogously.  As a notational shorthand, we often use 
$\p^\smallprup$ to represent
$\pru\delp{2}$;  we stress that both notations 
are used here to represent the class of sets accepted via {\em guardedly 
unambiguous\/} access to an NP oracle (that is, the class
of sets accepted by some P machine with an NP machine's language 
as its oracle
such that on no input does the P machine ask its oracle machine
any question on which the oracle machine has more than one accepting
path).

\item For each of the above hierarchies, we use 
${\rm \Sigma}_k^{p,A}$ (respectively, $\u{\rm \Sigma}_k^{p,A}$ and
$\pru{\rm \Sigma}_k^{p,A}$) 
to denote that the $\sigp{k}$
(respectively, $\u{\rm \Sigma}_k^{p}$ and
$\pru{\rm \Sigma}_k^{p}$) 
computation is performed relative to oracle $A$; similar notation
is used for the $\Pi$ and $\Delta$ classes of the hierarchies.
\end{enumerate}
\end{definition}

The following facts follow from the 
definition (see also~\cite{nie-ros:c:up-hierarchy}) or can easily be shown.

{\samepage
\begin{fact}
\quad
For every $k\geq 1$, 
\begin{enumerate}
\item $\u\sigp{k} \seq \pru\sigp{k} \seq \sigp{k}$ and 
$\u\delp{k} \seq \pru\delp{k} \seq \delp{k}$.

\item If $\u\sigp{k} = \u\pip{k}$, then $\u\ph = \u\sigp{k}$.

\item If $\u\sigp{k} = \u\sigp{k-1}$, then $\u\ph = \u\sigp{k-1}$.

\item $\u{\Sigma}_{k}^{p, {\protect\scriptsize \up \cap \coup}} = \u\sigp{k}$
  and $\p^{\protect\scriptsize \u\sigp{k} \cap \u\pip{k}} =
  \u\sigp{k}\cap\u\pip{k}$.
\end{enumerate}
\end{fact}
} %

The classes
``$\up_{\leq k}$,'' the analogs of UP in which up to $k$ accepting
paths are allowed, have been studied in various
contexts~\cite{wat:j:hardness-one-way,hem:thesis:counting,bei:c:up-additional,cai-hem-vys:b:promise,ehem-hem:j:quasi-injective,hem-zim:t:balanced-immunity}.
One motivation for $\u\sigp{k}$ is that, for each~$k$, $\up_{\leq k}
\seq \u\sigp{k}$~\cite{nie-ros:c:up-hierarchy}.

Although we are not able to settle affirmatively the question posed at
the end of the first paragraph of this section, we do prove in the
theorem below that if there is a sparse Turing-complete set for UP,
then the levels of the unambiguous polynomial hierarchy are simpler
than one would otherwise expect: they ``slip down'' slightly in terms
of their location within the promise unambiguous polynomial hierarchy,
i.e., for each $k\geq 3$, the $k$th level of UPH is contained in the
$(k-1)$st level of \pruph\@.

\begin{thm}\label{t:suitcase}
\quad
If there exists a sparse Turing-complete set for \up, then
\begin{enumerate}
\item $\up^{\protect\scriptsize \up} \seq \p^{\smallprup}$, and

\item $\u\sigp{k}\seq \pru\sigp{k-1}$ for every $k\geq 3$.
\end{enumerate}
\end{thm}

\noindent
{\bf Proof.} 
\quad
For the first statement, let $L$ be any set in $\up^{\protect\scriptsize
  \up}$. By assumption, $L\in \up^{\protect\scriptsize \p^{S}} = \up^{S}$ for
some sparse set $S\in \up$.  Let $q$ be a polynomial bounding the
density of $S$, that is, $\|S^{\leq m}\|\leq q(m)$ for every $m\geq
0$, and let $N_S$ be a UPM for $S$.  Let $N_L$ be a UPOM witnessing
that $L\in \up^{S}$, that is, $L=L(N_{L}^{S})$.  
Let $p(n)$ be a polynomial bounding the
length of all query strings that can be asked during the computation
of $N_L$ on inputs of length $n$. Define the polynomial 
$r(n) \equalsdef q(p(n))$ that bounds the number of strings 
in $S$ that can be queried in the run of $N_L$ on inputs of length $n$. 

To show that $L\in \p^\smallprup$, we shall construct a DPOM $M$ that
may access its \prup\ oracle $D$ in a guarded manner (more formally,
``may access its NP oracle $D$ in a guardedly unambiguous manner,''
but we will henceforward use $\cal UP$ and other $\cal U\cdots$
notations in this informal manner).  Before formally describing
machine $M$ (Figure~\ref{figure:DPOM}), we give some
informal explanations. $M$ will proceed in three basic steps: First,
$M$ determines the exact census of that part of $S$ that is relevant
for the given input length, $\|S^{\leq p(n)}\|$.  Knowing the exact
census, $M$ can construct (by prefix search) a table~$T$ of all
strings in $S^{\leq p(n)}$ without asking queries that make its
oracle's machine ambiguous, so the $\p^\smallprup$-like behavior is
guaranteed.  Finally, $M$ asks its oracle $D$ to simulate the
computation of~$N_L$ on input~$x$ (answering $N_L$'s oracle queries by
table-lookup using table~$T$), and accepts accordingly.

In the formal description of machine $M$ (given in
Figure~\ref{figure:DPOM}), three oracle sets $A$, $B$, and $C$ are
used.  Since $M$ has only one \tweak\prup\ oracle, the actual set to
be used is $D = A\oplus B\oplus C$ (with suitably modified queries to
$D$).  $A$, $B$, and $C$ are defined as follows (we assume the set $T$
below is coded in some standard reasonable way):
\begin{eqnarray*}
A & \equalsdef & \left\{\<1^n,k\>\
\begin{tabular*}{9.3cm}{|l}  %
$n\geq 0\,\wedge\, 0\leq k\leq r(n)\,\wedge\, 
(\exists c_1<_{\mbox{\protect\scriptsize lex}} c_2 <_{\mbox{\protect\scriptsize lex}} 
\cdots <_{\mbox{\protect\scriptsize lex}} c_k)$\\
$(\forall {\ell}:1\leq \ell \leq k)\,[|c_{\ell}|\leq p(n)\,\wedge\, 
N_{S}(c_{\ell})\ \mbox{accepts}\,]$
\end{tabular*}
\right\},
\\
B & \equalsdef & \left\{\<1^n,i,j,k,b\>\
\begin{tabular*}{8.8cm}{|l}
$n\geq 0\,\wedge\, 1\leq j\leq k\,\wedge\, 0\leq k\leq r(n)\,\wedge$\\
$(\exists c_1 <_{\mbox{\protect\scriptsize lex}} c_2 <_{\mbox{\protect\scriptsize lex}} 
\cdots <_{\mbox{\protect\scriptsize lex}} c_k)\, 
(\forall {\ell}:1\leq {\ell}\leq k)$\\ 
$[|c_{\ell}|\leq p(n) \,\wedge\, N_{S}(c_{\ell})\ \mbox{accepts}\,\wedge\, 
\mbox{the $i^{\mbox{\protect\scriptsize th}}$ bit of $c_j$ is $b$}\,]$
\end{tabular*}
\right\},
\\
C & \equalsdef & \{\<x,T\> \, | \,
\|T\|\leq r(|x|)\,\wedge\, N_{L}^{T}(x) \mbox{ accepts }\}.
\end{eqnarray*}

\begin{figure}

{ \singlespacing 
\begin{construction}  
  \item {\bf Description of DPOM} {\boldmath $M.$} 
  \begin{block}
    \item {\bf input} $x$; %
    \item {\bf begin} 
    \begin{block} 
      \item $n:= |x|$;
      \item $k:=r(n)$;
      \item {\bf loop}
      \begin{block}
        \item {\bf if} $\<1^{n},k\>\in A$ {\bf then exit loop} 
        \item {\bf else} $k:=k-1$
      \end{block}
      \item {\bf end loop}
                \hfill (* $k$ is now the exact census of $S^{\leq p(n)}$ *)
      \item $T:=\emptyset$;     \hfill (* $T$ collects the strings of
                                $S^{\leq p(n)}$ *)
      \item {\bf for} $j=1$ {\bf to} $k$ {\bf do}
        \begin{block}
          \item $c_j:=\epsilon$; 
          \item $i:=1$;
          \item {\bf repeat}
            \begin{block}
              \item {\bf if} $\<1^{n},i,j,k,0\>\in B$ 
{\bf then}
                        $c_j:=c_j0$; $i:=i+1$
              \item {\bf else} 
                \begin{block}
                  \item {\bf if} $\<1^{n},i,j,k,1\>\in B$ {\bf then}
                        $c_j:=c_j1$; $i:=i+1$
                  \item {\bf else} $i:=0$
                                \hfill (* the lex. $j^{\mbox{\protect\scriptsize th}}$
                                string of $S^{\leq p(n)}$ has no 
                                $i^{\mbox{\protect\scriptsize th}}$ bit *)
                \end{block}
            \end{block}
          \item {\bf until} $i=0$;
          \item $T:=T\cup \{c_j\}$
        \end{block}
      \item {\bf end for}
      \item {\bf if} $\<x,T\>\in C$ {\bf then accept}
      \item {\bf else reject}
    \end{block}
    \item {\bf end}
  \end{block}
  \item {\bf End of Description of DPOM} {\boldmath $M.$}
\end{construction}

} %
\caption{\label{figure:DPOM} DPOM $M$ guardedly unambiguously accessing 
an NP oracle 
to accept a set in $\up^{\protect\scriptsize \up}$.}
\end{figure}

It is easy to see that $M$ runs deterministically in polynomial time.
This proves that $L\in \p^\smallprup$.

\smallskip

In order to prove the second statement, let $L$ be a set in
$\u\sigp{k}$ for any fixed $k\geq 3$. By assumption, there exists a
sparse set $S$ in \up\ such that $L\in \u{\Sigma}_{k-1}^{p,
  {\protect\scriptsize \p}^{S}} = \u{\Sigma}_{k-1}^{p,S}$; let
$N_1, N_2, \ldots,N_{k-1}$ be the UPOMs that witness this fact, that is, $L
= L(N_{1}^{L(N_{2}^{\.^{\.^{L(N^S_{k-1})}}})})$.

Now we describe the computation of a $\pru\sigp{k-1}$ machine $N$
recognizing~$L$. As before, $N$~on input~$x$ computes in
$\p^\smallprup$ its table of advice strings, \mbox{$T = S^{\leq
    p(|x|)}$}, and then simulates the $\u{\Sigma}_{k-1}^{p,S}$
computation of 
$N_{1}^{L(N_{2}^{\.^{\.^{L(N^S_{k-1})}}})}(x)$ 
except with $N_1$, $N_2$, $\ldots$, 
$N_{k-1}$ modified as follows.
If in the simulation some
machine $N_i$, $1 \leq i\leq k-2$, consults its original oracle
$L(N_{i+1}^{(\cdot )})$ about some string, say~$z$, then the modified
machine $N_{i}^{'}$ queries the modified machine at the next level,
$N_{i+1}^{'}$, about the string $\<z,T\>$ instead. Finally, the advice
table $T$, which has been ``passed up'' in this manner, is used to
correctly answer all queries of $N_{k-1}$.

Note that $N$'s oracle in this simulation,
$L({N_{2}^{'}}^{L({N_{3}^{'}}^{\.^{\.^{L(N_{k-1}^{'})}}})})$, is not
in general a $\u\sigp{k-2}$ set (and~$L$ is thus not in $\u\sigp{k-1}$
in general), as the above-described computation depends on the advice
table~$T$, and so, for some bad advice $T$, the unambiguity of the
modified machines $N_{1}^{'}, N_{2}^{'}, \ldots , 
N_{k-1}^{'}$ is no longer
guaranteed.  But since our base machine $N$ is able to provide {\em
  correct\/} advice~$T$, we have indeed shown that $L\in
\pru\sigp{k-1}$.~\hfill$\Box$

\medskip

In the above proof, the assumption that the sparse set $S$ is in \up\
is needed to determine the exact census of $S$ using the UPM for $S$.
Let us now consider the weaker assumption that UP has only a 
Turing-{\em hard\/} sparse set.  Karp and Lipton have shown that if there is
a sparse Turing-hard set for \np, then the polynomial hierarchy
collapses to its second level~\cite{kar-lip:c:nonuniform}.\footnote{%
\protect\singlespacing
Very recently, K\"obler and 
Watanabe~\protect\cite{koe-wat:t:new-collapse} have improved this 
collapse to 
$\mbox{ZPP}^{\tiny \np}$, and have also obtained new consequences from the 
assumption that $\up \seq (\np \cap \conp)/\mbox{poly}$, whereas 
we obtain different consequences from the assumption that
$\up \seq \p/\mbox{poly}$ (see~\protect\cite{koe-wat:t:new-collapse} for
the notations not defined in this footnote).}
Hopcroft~\cite{hop:c:recent} dramatically simplified their proof, and
Balc\'{a}zar, Book, and
Sch\"oning~\cite{bal-boo-sch:j:sparse,sch:b:complexity-structure}
generalized, as Theorem~\ref{t:tee}, the Karp-Lipton result; the
general approach of Hopcroft and Balc\'{a}zar, Book, and Sch\"oning
will be central to our upcoming proof of Theorem~\ref{t:glass}.
Sch\"oning's low hierarchy~\cite{sch:j:low} gives a way of classifying
the complexity of NP sets that seem to be neither in P nor NP-complete. Of
particular interest to us is the class \mbox{$\mbox{Low}_2 \equalsdef
  \{ A \,|\, A \in \np\ \mbox{and}\ \np^{\protect\scriptsize \np^{A}}
    \seq \np^{\protect\scriptsize \np} \}$}. Note that for the special
  case $k = 0$, Theorem~\ref{t:tee} below says that 
  $\mbox{Low}_2 \supseteq \np  \cap  {\rm P}/{\rm poly} 
  \cap \{ L \, | \, L$ is self-reducible$\}$.

\begin{definition} \cite{mey-pat:t:int}~
\begin{enumerate}
\item A partial order $<_{\mbox{\protect\scriptsize pwl}}$ on $\Sigma^*$ is
  {\em polynomially well-founded and length-related\/} if and only if
  (a)~every strictly decreasing chain is finite and there is a
  polynomial $p$ such that every finite $<_{\mbox{\protect\scriptsize
      pwl}}$-decreasing chain is shorter than $p$ of the length of its
  maximum element, and (b)~$(\exists q: q\ 
  \mbox{polynomial})\,(\forall x,y\in\Sigma^*)\,
  [x<_{\mbox{\protect\scriptsize pwl}}y \Lora |x|\leq q(|y|)]$.

\item A set $A$ is {\em self-reducible\/} if and only if there exist
  a polynomially well-founded and length-related order
  $<_{\mbox{\protect\scriptsize pwl}}$ on $\Sigma^*$ and a DPOM $M$ such that
  $A=L(M^A)$ and on any input $x\in \Sigma^*$, $M$ queries only
  strings $y$ with $y<_{\mbox{\protect\scriptsize pwl}}x$.
\end{enumerate}
\end{definition}

\begin{lem}\label{l:golf}
\cite{bal-boo-sch:j:sparse}
\quad
Let $A$ be a self-reducible set and let $M$ witness $A$'s 
self-reducibility. 
For any set $B$ and any $n$, if $\left(L(M^B)\right)^{\leq n}=B^{\leq n}$, 
then $A^{\leq n}=B^{\leq n}$\@.\footnote{%
\protect\singlespacing
$A$ can be viewed as a ``fixed point'' of $M$.
}
\end{lem}

\begin{thm} \label{t:tee}
\cite{bal-boo-sch:j:sparse}
\quad
If $A$ is a self-reducible set and there is a $k\geq 0$ and a sparse 
set $S$ such that $A\in {\Sigma}_{k}^{p,S}$, 
then ${\Sigma}_{2}^{p,A}\seq \sigp{k+2}$.
\end{thm}

We now state and prove our results regarding sparse Turing-hard sets
for UP\@.

\begin{thm}\label{t:glass}
\quad
If there exists a sparse Turing-hard set for \up, then
\begin{enumerate}
\item $\up \seq \mbox{Low}_2$, and 

\item
$\u\sigp{k} \seq 
  \u{\Sigma}_{j}^{p,{\textstyle 
\Sigma}_{2}^{p, \tweak\tinypru{\Sigma}_{k-j-3}^p}}
  \cap \p^{\smallpru\sigp{k-1}{\oplus} \sigp{2}}$
\ for every $k\geq 3$ and every $j$, with $0\leq j\leq k-3$.
\end{enumerate}
\end{thm}

\noindent 
{\bf Proof.} 
\quad
1. Let $L\in {\Sigma}_{2}^{p,A}$, where $A\in \up$ via UPM $N_A$ and
polynomial-time bound $t$ (we assume that each step is
nondeterministic---one can require this, without loss of
generality, while maintaining categoricity). 
Our proof uses the well-known fact that the ``left set''
\cite{sel:j:natural,ogi-wat:j:sparse-btt-complete} of any UP set is
self-reducible and is in UP\@.  More precisely, to apply
Theorem~\ref{t:tee} we would need~$A$ to be self-reducible. Although
that can't be assumed in general of an arbitrary \up\ set, the left
set of~$A$, i.e., the set of prefixes of witnesses for elements in $A$
defined by
$$B \equalsdef
\{\<x,y\>\,|\,(\exists z)\,[|yz|=t(|x|)\,\wedge\, N_{A}(x)\ 
\mbox{accepts on path}\ yz]\},$$ 
does have this property and is also in \up\@.  A self-reducing machine
$M_{\mbox{\em \scriptsize self}}$ for $B$ is given in
Figure~\ref{figure:self-reducer}.  Note that the queries asked in the
self-reduction are strictly less than the input with respect to a
polynomially well-founded and length-related partial order
$<_{\mbox{\protect\scriptsize pwl}}$ defined by: For fixed $x$ and all strings
$y_1,y_2\in {\Sigma}^{\leq p(|x|)}$, $\<x,y_1\><_{\mbox{\protect\scriptsize
    pwl}}\<x,y_2\>$ if and only if $y_2$ is prefix of $y_1$.  

By assumption, since $B$ is a UP set, $B \in \p^S$ for some sparse set
$S$, so Theorem~\ref{t:tee} with $k=0$ applies to $B$. Furthermore,
$A$ is in $\p^B$, via prefix search by DPOM $M_A$
(Figure~\ref{figure:turing}).  Thus, $L\in {\Sigma}_{2}^{p,
  {\protect\scriptsize \p}^{B}}\seq {\Sigma}_{2}^{p,B} \seq \sigp{2}$, which
shows that $A \in \mbox{Low}_2$.

\smallskip

\begin{figure}
\begin{center}
\parbox{8cm}
{\singlespacing 

\begin{construction}  
\item {\bf Description of Self-reducer} 
{\boldmath $M_{\mbox{\protect\scriptsize\bf self}}$} {\bf for}
  {\boldmath $B.$}
  \begin{block}
    \item {\bf input} $\<x,y\>$;
    \item {\bf begin} 
    \begin{block} 
      \item {\bf if} $|y| > t(|x|)$ {\bf then reject};
      \item {\bf if} $N_{A}(x)$ accepts on path $y$ {\bf then accept}
      \item {\bf else}  
      \begin{block}
        \item {\bf if} $\<x,y0\>\in B$ or $\<x,y1\>\in B$ {\bf then accept} 
        \item {\bf else reject} 
      \end{block}
    \end{block}
    \item {\bf end}
  \end{block}
\item {\bf End of Description of Self-reducer} 
{\boldmath $M_{\mbox{\protect\scriptsize\bf self}}$}
  {\bf for} {\boldmath $B.$}
\end{construction}

} %
\end{center}
\caption{\label{figure:self-reducer} A self-reducing machine for the 
left set of a UP set.}
\end{figure}

\begin{figure}
\begin{center}
\parbox{6cm}
{\singlespacing

\begin{construction}  
  \item {\bf Description of DPOM} {\boldmath $M_A.$} 
  \begin{block}
    \item {\bf input} $x$;
    \item {\bf begin} 
    \begin{block} 
      \item $y:=\epsilon$;
      \item {\bf while} $|y|<t(|x|)$ {\bf do}
      \begin{block}
        \item {\bf if} $\<x,y0\>\in B$ {\bf then accept} 
        \item {\bf else} $y:=y1$
      \end{block}
      \item {\bf end while}
      \item {\bf if} $\<x,y\>\in B$ {\bf then accept}
      \item {\bf else reject}  
    \end{block}
    \item {\bf end}
  \end{block}
  \item {\bf End of Description of DPOM} {\boldmath $M_A.$}
\end{construction}

} %
\end{center}
\caption{\label{figure:turing} A Turing reduction 
from a UP set $A$ to its left set $B$ via prefix search.}
\end{figure}

2. For $k=3$ (thus $j=0$), both inclusions have already been shown in
Part~1, as $\sigp{2}\seq \delp{3}$.  Now fix any $k>3$, and let $L\in
\u\sigp{k}=\u{\Sigma}_{k-1}^{p,A}$ be witnessed by UPOMs
$N_1, N_2, \ldots ,N_{k-1}$ and \mbox{$A\in \up$}.  
Define $B$ to be the left set of $A$ as in
Part 1, so $A\in \p^B$ via DPOM $M_A$ (see
Figure~\ref{figure:turing}), $B$ is self-reducible via
$M_{\mbox{\em \scriptsize self}}$ (see
Figure~\ref{figure:self-reducer}), and $B$ is in \up\@. By hypothesis,
$B\in \p^S$ for some sparse set~$S$; let $M_B$ be the reducing
machine, that is $B = L(M_{B}^{S})$, and let $m$ be a polynomial bound
on the runtime of~$M_B$.  Let $q$ be a polynomial such that $\|S^{\leq
  m}\|\leq q(m)$ for every $m\geq 0$.  Let $p(n)$ be a polynomial
bounding the length of all query strings whose membership in the
oracle set $B$ can be asked in the run of $N_1$ (with oracle machines
$N_2$, $N_3$, $\ldots$, $N_{k-1}$, $M_{A}^{B}$) on inputs of length~$n$.  
Define the polynomials $r(n) \equalsdef m(p(n))$ and
$s(n) \equalsdef q(r(n))$.

To show that $L\in \p^{\smallpru\sigp{k-1}{\oplus} \sigp{2}}$, we will
describe a DPOM $M$ that on input $x$, $|x| = n$, using the $\sigp{2}$
part $D$ (defined below) of its oracle, performs a prefix search to
extract the lexicographically smallest of all ``good'' advice sets 
(this informal term will be formally defined in the next paragraph),
say $T$, and then calls the $\pru\sigp{k-1}$ part of its oracle to 
simulate the $\u{\Sigma}_{k-1}^{p,A}$ computation of 
$N_{1}^{L(N_{2}^{\.^{\.^{L(N^A_{k-1})}}})}(x)$ 
except with $N_1$, $N_2$, $\ldots$, 
$N_{k-1}$ 
modified in the same way as was described in the proof 
of Theorem~\ref{t:suitcase}. In more detail, if in the simulation some
machine $N_i$, $1 \leq i\leq k-2$, consults its original oracle
$L(N_{i+1}^{(\cdot )})$ about some string, say~$z$, then the modified
machine $N_{i}^{'}$ queries the modified machine at the next level,
$N_{i+1}^{'}$, about the string $\<z,T\>$ instead. Finally, if
$N_{k-1}$ consults its original oracle $A$ about some query~$y$, then the
modified machine $N_{k-1}^{'}$ runs the P computation
$M_{A}^{L(M_{B}^{T})}$ on input $\<y,T\>$ instead to
correctly answer this query without consulting an oracle.  

An advice set $T$ is said
to be {\em good\/} if the set $L(M_{B}^{T})$ is a fixed point of $B$'s
self-reducer $M_{\mbox{\em \scriptsize self}}$ up to length $p(n)$,
that is, $\left( L(M_{\mbox{\em \scriptsize self}}^{L(M_{B}^{T})})
\right)^{\leq p(n)} = \left( L(M_{B}^{T})\right)^{\leq p(n)}$, and
thus $B^{\leq p(n)} = \left(L(M_{B}^{T})\right)^{\leq p(n)}$ by
Lemma~\ref{l:golf}.  This property is checked for each guessed $T$ in
the $\sigp{2}$ part of the oracle. Formally,
\[
D \equalsdef \left\{\<1^n,i,j,b\>\
\begin{tabular*}{10.3cm}{|l}
$n\geq 0\,\wedge\, (\exists T \seq \Sigma^{\leq r(n)})\,
(\forall w : |w| \leq p(n)\,)\,
[ T = \{c_1,\ldots ,c_k\}$ \\
$\wedge\, 0\leq k\leq s(n) \,\wedge\, c_1 <_{\mbox{\protect\scriptsize lex}} 
\cdots <_{\mbox{\protect\scriptsize lex}} c_k \,\wedge\, 
\mbox{the $i^{\mbox{\protect\scriptsize th}}$ bit of $c_j$ is $b$ } \,\wedge$ \\
$(w\in L(M_{B}^{T})\Lolra w\in L(M_{\mbox{\em \scriptsize self}}^{L(M_{B}^{T})}))]$
\end{tabular*}
\right\}.
\]
The prefix search of $M$ is similar to the one performed in the proof
of Theorem~\ref{t:suitcase} (see Figure~\ref{figure:DPOM});
$M$ queries $D$ to construct each string of $T$ bit by bit.

To prove the other inclusion, fix any $j$, $0\leq j\leq k-3$.
We describe a UPOM $N$ witnessing that
$L\in \u{\Sigma}_{j}^{p,{\textstyle \Sigma}_{2}^{p, \tweak\tinypru
\sigp{k-j-3}}}$.
On input $x$, $N$ simulates the $\u\sigp{j}$ computation of the
first $j$ \mbox{UPOMs} $N_1,\ldots ,N_j$. 
In the subsequent $\sigp{2}$ computation, two tasks 
have to be solved in parallel: 
the computation of $N_{j+1}$ and $N_{j+2}$ is to be simulated, 
and good advice sets $T$ have to be determined. 
For the latter task, the base machine of the $\sigp{2}$ computation
guesses all possible advice sets and the top machine checks if the
guessed advice is good (that is, if $L(M_{B}^{T})$ is a fixed point of 
$M_{\mbox{\em \scriptsize self}}$).
Again, each good advice set $T$ is ``passed up'' to the machines at
higher levels
\mbox{$N_{j+3},\ldots ,N_{k-1}$} (in the same fashion as was employed 
earlier in this proof and also in the proof of 
Theorem~\ref{t:suitcase}), and is used to correctly answer all queries of
$N_{k-1}$ without consulting an oracle. This proves the theorem.~\hfill$\Box$

\medskip

Since Theorem~\ref{t:glass} relativizes and there are relativized
worlds in which $\up^A$ is not
$\mbox{Low}_{2}^{A}$~\cite{she-lon:tOUTDATED:up-low-relativized}, we
have the following corollary.

\begin{corollary}
\label{cor:glass} 
There is a relativized world in which (relativized) UP has no sparse
Turing-hard sets. 
\end{corollary}

{\samepage
\begin{center}
{\bf Acknowledgments}
\end{center}
\nopagebreak
\indent
We are very grateful to Gerd Wechsung for his help in bringing about
this collaboration, and for his kind and insightful advice over many
years.  We thank Marius Zimand for proofreading, and Nikolai
Vereshchagin for helpful discussions during his visit to Rochester.
We thank Osamu Watanabe for 
discussing with us his results joint with Johannes K\"obler, and we
thank Osamu Watanabe and Johannes K\"obler for providing us with
copies of their paper~\cite{koe-wat:t:new-collapse}.

}%

{\singlespacing

{\bibliography{main.bbl}}
}

\end{document}